\documentclass[11pt,twoside]{article}
\usepackage{./asp2014}

\def \der{{\rm d}}
\usepackage{color}
\aspSuppressVolSlug
\resetcounters

\begin{document}

\title{HI intensity mapping with FAST}
\author{M.-A.~Bigot-Sazy$^{\dagger}$, Y.-Z.~Ma$^{\star,\dagger}$, R.~A.~Battye$^{\dagger}$, I.~W.~A.~Browne$^{\dagger}$, T.~Chen$^{\dagger}$, C.~Dickinson$^{\dagger}$, S.~Harper$^{\dagger}$, B.~Maffei$^{\dagger}$, L.~C.~Olivari$^{\dagger}$, P.~N.~Wilkinson$^{\dagger}$}
\affil{$^{\dagger}$Jodrell Bank Centre for Astrophysics, School of Physics and Astronomy, The University of Manchester, Oxford Road, Manchester M13 9PL, UK} 
\affil{$^{\star}$(Corresponding author)\,School of Chemistry and Physics, University of KwaZulu-Natal, Westville Campus, Private Bag X54001, Durban, 4000, South Africa;\email{ma@ukzn.ac.za}}


\paperauthor{Marie-Anne Bigot-Sazy}{marie-anne.bigot-sazy@manchester.ac.uk}{}{University of Manchester}{Jodrell Bank Centre for Astrophysics}{Manchester}{Lancashire}{M13 9PL}{United Kingdom}
\paperauthor{Yin-Zhe Ma}{ma@ukzn.ac.za}{}{University of KwaZulu-Natal}{School of Chemistry and Physics}{Durban}{KwaZulu-Natal}{X54001}{South Africa}
\paperauthor{Richard Battye}{Richard.Battye@manchester.ac.uk}{}{University of Manchester}{Jodrell Bank Centre for Astrophysics}{Manchester}{Lancashire}{M13 9PL}{United Kingdom}
\paperauthor{Ian Browne}{ian.browne@manchester.ac.uk}{}{University of Manchester}{Jodrell Bank Centre for Astrophysics}{Manchester}{Lancashire}{M13 9PL}{United Kingdom}
\paperauthor{Tianyue Chen}{tianyue.chen@postgrad.manchester.ac.uk}{}{University of Manchester}{Jodrell Bank Centre for Astrophysics}{Manchester}{Lancashire}{M13 9PL}{United Kingdom}
\paperauthor{Clive Dickinson}{clive.dickinson@manchester.ac.uk}{}{University of Manchester}{Jodrell Bank Centre for Astrophysics}{Manchester}{Lancashire}{M13 9PL}{United Kingdom}
\paperauthor{Stuart Harper}{stuart.harper@manchester.ac.uk}{}{University of Manchester}{Jodrell Bank Centre for Astrophysics}{Manchester}{Lancashire}{M13 9PL}{United Kingdom}
\paperauthor{Bruno Maffei}{Bruno.Maffei@manchester.ac.uk}{}{University of Manchester}{Jodrell Bank Centre for Astrophysics}{Manchester}{Lancashire}{M13 9PL}{United Kingdom}
\paperauthor{Lucas Collis Olivari}{lucas.olivari@postgrad.manchester.ac.uk}{}{University of Manchester}{Jodrell Bank Centre for Astrophysics}{Manchester}{Lancashire}{M13 9PL}{United Kingdom}
\paperauthor{Peter Wilkinson}{peter.wilkinson@manchester.ac.uk}{}{University of Manchester}{Jodrell Bank Centre for Astrophysics}{Manchester}{Lancashire}{M13 9PL}{United Kingdom}

\begin{abstract}
We discuss the detectability of large-scale HI intensity fluctuations using the FAST telescope. We present forecasts for the accuracy of measuring the Baryonic Acoustic Oscillations and constraining the properties of dark energy. The FAST $19$-beam L-band receivers ($1.05$--$1.45$ GHz) can provide constraints on the matter power spectrum and dark energy equation of state parameters ($w_{0},w_{a}$) that are comparable to the BINGO and CHIME experiments. For one year of integration time we find that the optimal survey area is $6000\,{\rm deg}^2$. However, observing with larger frequency coverage at higher redshift ($0.95$--$1.35$\,GHz) improves the projected errorbars on the HI power spectrum by more than $2~\sigma$ confidence level. The combined constraints from FAST, CHIME, BINGO and {\it Planck} CMB observations can provide reliable, stringent constraints on the dark energy equation of state.

\end{abstract}

\section{Introduction}
\label{sec:intro}

Cosmology has been moving towards a ``precision'' era during which the cosmological parameters have been measured to high accuracy. In particular, the recent development in the observations of cosmic microwave background radiation~\citep{Planck-13}, galaxy spectroscopic and photometric surveys (e.g. SDSS-III~\citep{Reid14}, and Dark Energy Survey~\citep{Abbott15}) have allowed us to survey progressively large volumes of the Universe and thus more and more modes of fluctuations. The fundamental limit on the measurement accuracy of the primordial fluctuations is set by the volume of the observable Universe, known as cosmic variance. Thus the best we can do is to map out the entire observable volume of the Universe. 

Atomic hydrogen (HI) is expected to be a good tracer of the underlying dark matter distribution on large-scales and therefore an excellent indicator of structure formation~(e.g.~\citealt{Santos15}). The 21-cm emission from hyperfine structure splitting of HI is emitted at a fixed emission frequency (1420\,MHz), therefore measurement of its wavelength can reveal its emission redshift which can be used to reconstruct a 3-dimensional picture of cosmic structures.

The bulk of the HI signal is thought to come from very dense gas clouds that are shielded from ionised UV photons~\citep{Santos15}. However, 21-cm emission from individual galaxies is very weak. By observing with a beam large enough to integrate together the emission of many galaxies the sensitivity requirements of any given observation can be relaxed. Little cosmological information is lost by lost by smoothing out the small-scale structures of collapsed objects. On large-scales (approximately $1$-degree at $z<3$) baryonic acoustic oscillations (BAO) imprints a characteristic scale of $\sim150\,h^{-1}{\rm Mpc}$ within the HI matter distribution. Measurements of the  matter power spectrum and the BAO signal can be used to constrain cosmological parameters and probe the nature of dark energy~\citep{Battye13}.

The Five-hundred-meter Aperture Spherical radio Telescope (FAST) is a multi-beam radio telescope under construction in the south west of China~\citep{Nan11,Smoot14}. It covers a large frequency range of $70$\,MHz--$3$\,GHz and an area of sky that can potentially be used for 21-cm IM observation. Previously, \citet{Smoot14} have provided forecasts on how well FAST can measure the matter power spectrum between redshifts $0.5$--$2.5$. However, the authors assume that the total aperture is 500 metres while in reality the illuminated aperture is 300 metres~\citep{Nan11}. They also assume that the complete set of receivers proposed by~\citet{Nan11} can be built and placed in the focal plane array and used at the same time for a dedicated survey. We note that only a sub-set of these receivers will likely be useful for intensity mapping, especially since some of the receivers have only a single pixel. Furthermore, detectors away from the optical axis will suffer beam degradation. In this paper we will suggest an optimal scanning strategy for FAST, and present a detailed analysis of the capability of 21-cm IM with FAST given $1$-year observational time. In addition, we will compare its prospective cosmological constraints with other two on-going radio surveys: Canadian Hydrogen Intensity Mapping Experiment (CHIME) being built in western Canada~\citep{Shaw14} and BAO with Integrated Neutral Gas Observations (BINGO) proposed to be built in Uruguay~\citep{Battye12}.

This paper is organised as follows. In Section~\ref{sec:signal-noise}, we will discuss the 21-cm signal from HI gas, the measurement noise, the algorithm to be used to forecast the parameter precision, and also describe the experimental specification of FAST. We will present our main results in Section~\ref{sec:result}, including the optimal survey area, constraints on the BAO scale and dark energy equation of state (EoS). Concluding remarks will be presented in the last section. Throughout this paper, unless otherwise stated, we adopt a spatially-flat $\Lambda$ Cold-Dark-Matter cosmology model with best-fitting cosmological parameters measured with {\it Planck} 2015 results~\citep{Planck-13}: $\Omega_{\rm m}=0.308$, $\Omega_{\rm b}=0.048$, $\Omega_{\Lambda}=0.691$, $\sigma_{8}=0.834$, $n_{\rm s}=0.967$, $h=0.67$.

\section{21-cm signal, thermal noise, and FAST experimental parameters}
\label{sec:signal-noise}

\subsection{21-cm signal}
The mean temperature of observational 21-cm emission is~\citep{Battye13,Hall13}
\begin{eqnarray}
\overline{T}_{\rm b}(z) =188 \Omega_{\rm HI}(z)h \frac{(1+z)^2}{E(z)}{\rm mK} = 127 \left(\frac{h}{0.7} \right)\left(\frac{\Omega_{\rm HI}(z)}{10^{-3}} \right)\left(\frac{(1+z)^{2}}{E(z)} \right) {\rm \mu K},
 \label{eq:Tb}
\end{eqnarray}
where $\Omega_{\rm HI}(z)$ is the comoving mass density in HI relative to the present critical density, and $E(z)=\sqrt{\Omega_{\rm m}(1+z)^{3}+\Omega_{\Lambda}}$. 

\subsection{Thermal noise}
We calculate the projected errorbars on the resulting power spectra by using the 3D power spectrum (\ref{eq:3d-power}) as ~\citep{Seo10} 
\begin{eqnarray}
\frac{\sigma_{P}}{P}=\sqrt{2\frac{(2\pi)^{3}}{V_{\rm sur}}\frac{1}{4\pi k^{2}\Delta k}}\left(1+\frac{\sigma^{2}_{\rm pix}V_{\rm pix}}{\left[\bar{T}_{\rm b}(z) \right]^{2}W^{2}(k)P} \right), \label{eq:deltaP}
\end{eqnarray}
where
\begin{eqnarray}
V_{\rm sur}=\Omega_{\rm sur}\int^{z_{\rm max}}_{z_{\rm min}}\der z \left(\frac{\der^{2}V}{\der z \der \Omega} \right),\,\,V_{\rm pix}=\Omega_{\rm pix}\int^{z+\Delta z/2}_{z-\Delta z/2}\der z \left(\frac{\der^{2}V}{\der z \der \Omega} \right)
\end{eqnarray}
are the survey and pixel volume respectively, and $\der^{2}V/\der z \der \Omega=c\chi^{2}(z)/H(z)$ is the comoving volume per redshift per solid angle. \
\begin{eqnarray}
\sigma_{\rm pix}= \frac{T_{\rm sys}}{\sqrt{t_{\rm pix}\Delta f}},\,\, t_{\rm obs}=\frac{\Omega_{\rm sur}}{n_{\rm F}\Omega_{\rm pix}} t_{\rm pix},
\end{eqnarray}
are the r.m.s. noise of the pixel and the relation between total and individual pixel observational times, respectively. The former is related to the system temperature $T_{\rm sys}$, and the observational time per pixel $t_{\rm pix}$ and frequency band width $\Delta f$, while the later is related to total, $\Omega_{\rm sur}$, and pixel, $\Omega_{\rm pix}$, survey areas, and the number of feedhorns at the focus of the telescope, $n_{\rm F}$. For a fixed total observational time, the larger the sky coverage is, the smaller the observational time per pixel. This causes an increase of the r.m.s. noise per pixel, but it is compensated by a reduction in the cosmic variance. 

\begin{table}
\begin{centering}
\begin{tabular}{@{}c c c}\hline
& FAST (Extended) & BINGO \\ \hline
Frequency range [MHz] &  $950$--$1350$ &  $960$--$1260$ \\ \hline
Band width\footnote{This refers to single channel} [MHz] &  $1$ &  $1$ \\ \hline
Number of dishes &  $1$ &  $1$ \\ \hline
Number of beams &  $19$ &  $70$ \\ \hline
Single-dish diameter [m] &  $300$ &  $40$ \\ \hline
System temperature [K] &  $35$ &  $50$ \\ \hline
Survey area [deg$^{2}$]&   $6000$ & $3000$ \\ \hline
\noalign{\vspace{-1.5pt}} \hline
\end{tabular}%
\caption{The survey parameters of IM experiments using the FAST and BINGO instruments. Here we fix the total observational period to be $1$-year. For FAST, the total illuminated aperture is assumed to be $300$ metres~\citep{Nan11}.} \label{tab:surveys}
\end{centering}
\end{table}

\subsection{FAST experimental specification}
\label{sec:fast-experiment}
\begin{figure}[tbp]
\centerline{
\includegraphics[width=3.4in]{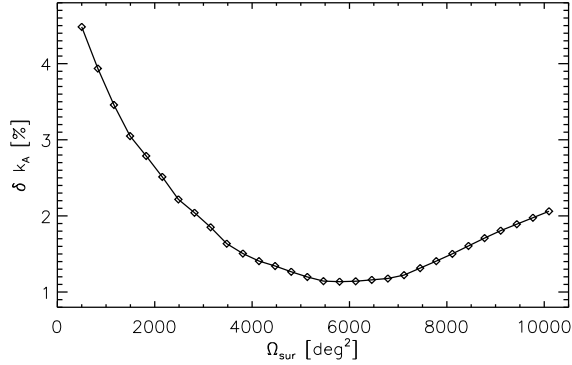}}
\caption{Projected error of the wavenumber of the BAO scale as a function of survey area $\Omega_{\rm sur}$ for a fixed observation time of $1$-year. The optimal survey area is $\simeq 6000\,$deg$^{2}$.} \label{fig:Omega_A}
\end{figure}

In Table~\ref{tab:surveys}, we list the basic experimental parameters for FAST and BINGO. In table 2 of~\citet{Nan11}, one can see that it is planned to have $19$ beams in one receiver array covering a frequency range (L-band) $1.05$--$1.35$\,GHz. The upper frequency limit of $1.35\,$GHz was chosen because at higher frequencies, corresponding to lower redshifts, the HI matter power spectrum and BAO signal are difficult to measure. In addition, the receiver front end can work at lower frequencies, i.e. higher redshift (Maffei 2015 private communication). We propose an extended survey for the FAST experiment with a larger frequency coverage $0.95$--$1.35$\,GHz. For the BINGO survey parameters, we consider the values defined in the paper of \citet{Battye12}. We use these parameters in the procedure of forecasting. In what follows, we will analyze the expected performance of the FAST telescope for measuring the BAO wiggles in the $k$ range $0$--$0.35$\,Mpc$^{-1}$. We will fix the total observational time to $1$-year and consider the survey parameters defined in ~\citet{Nan11}, except extending to larger frequency coverage.

The BAO fluctuations in the 21-cm matter power spectrum at each redshift ($\nu_{\rm obs}=1.42{\,\rm GHz}/(1+z)$) can be approximated by a sinusoidal function~\citep{Blake03,Battye13}
\begin{eqnarray}
P(k,z)=P_{\rm smo}(k,z)\left[1+A k \exp\left(-\left(\frac{k}{0.1\,h\,{\rm Mpc}^{-1}} \right)^{1.4} \right) \right]\sin\left(\frac{2 \pi k}{k_{A}} \right), \label{eq:3d-power} \label{eq:fit}
\end{eqnarray}
where $P_{\rm smo}(k,z)$ is the smoothed matter power spectrum at redshift $z$, $k_{A}$ is the wavenumber of the BAO scale. We then vary the survey area $\Omega_{\rm sur}$ to find the minimal error of BAO scale $k_{A}$. The result is shown in Fig.~\ref{fig:Omega_A}. One can see that the optimal collective survey area is at $6000\,$deg$^{2}$, which can be achieved by tilting the declination of the telescope each day. This optimal survey area is a balance of minimizing both the thermal noise and cosmic variance (Eq.~(\ref{eq:deltaP})).

\section{Results}
\label{sec:result}

\begin{figure}[tbp]
\centerline{\includegraphics[bb=0 0 600 300,width=3.2in]{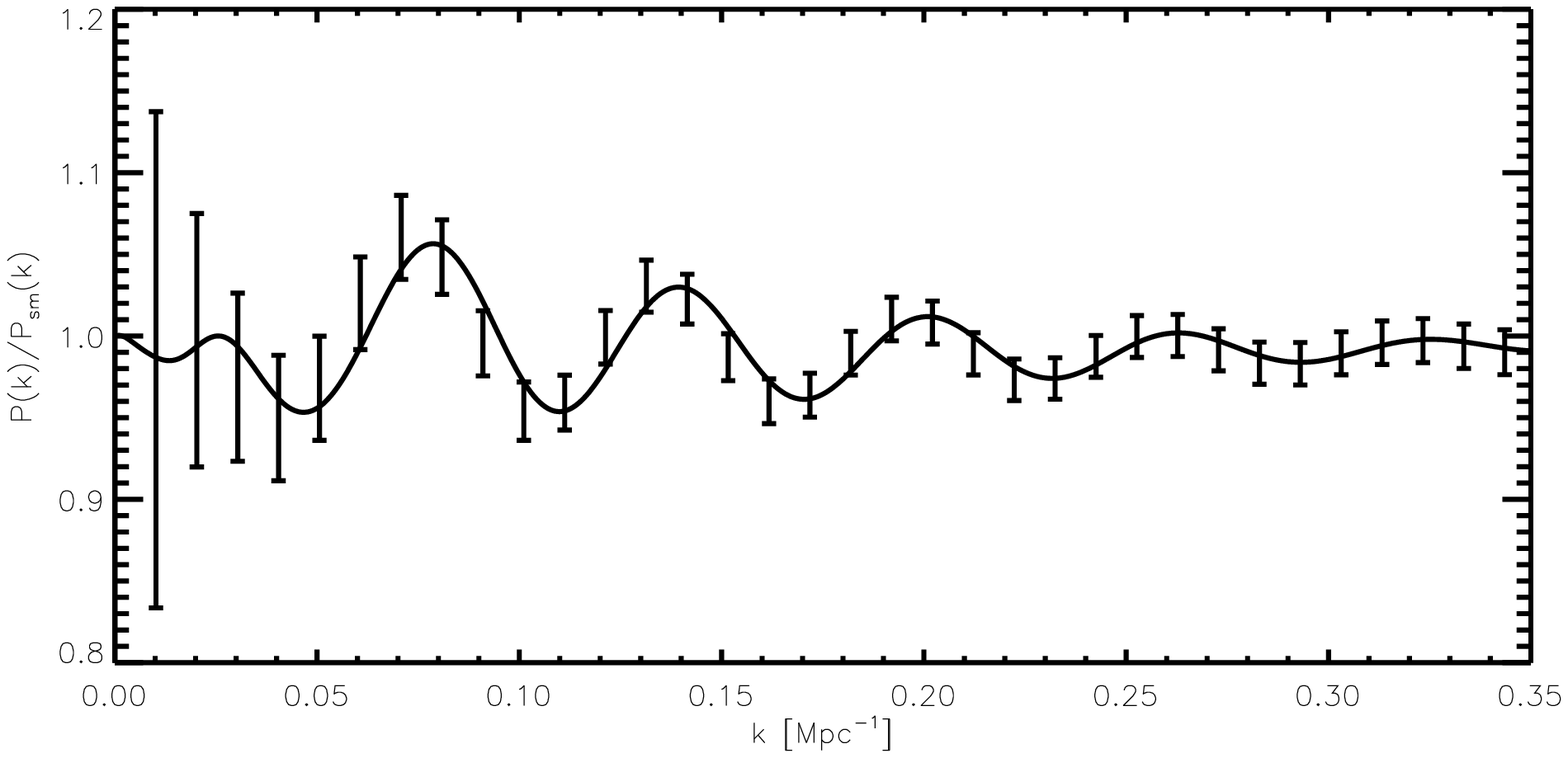}
\includegraphics[bb=0 0 600 300, width=3.2in]{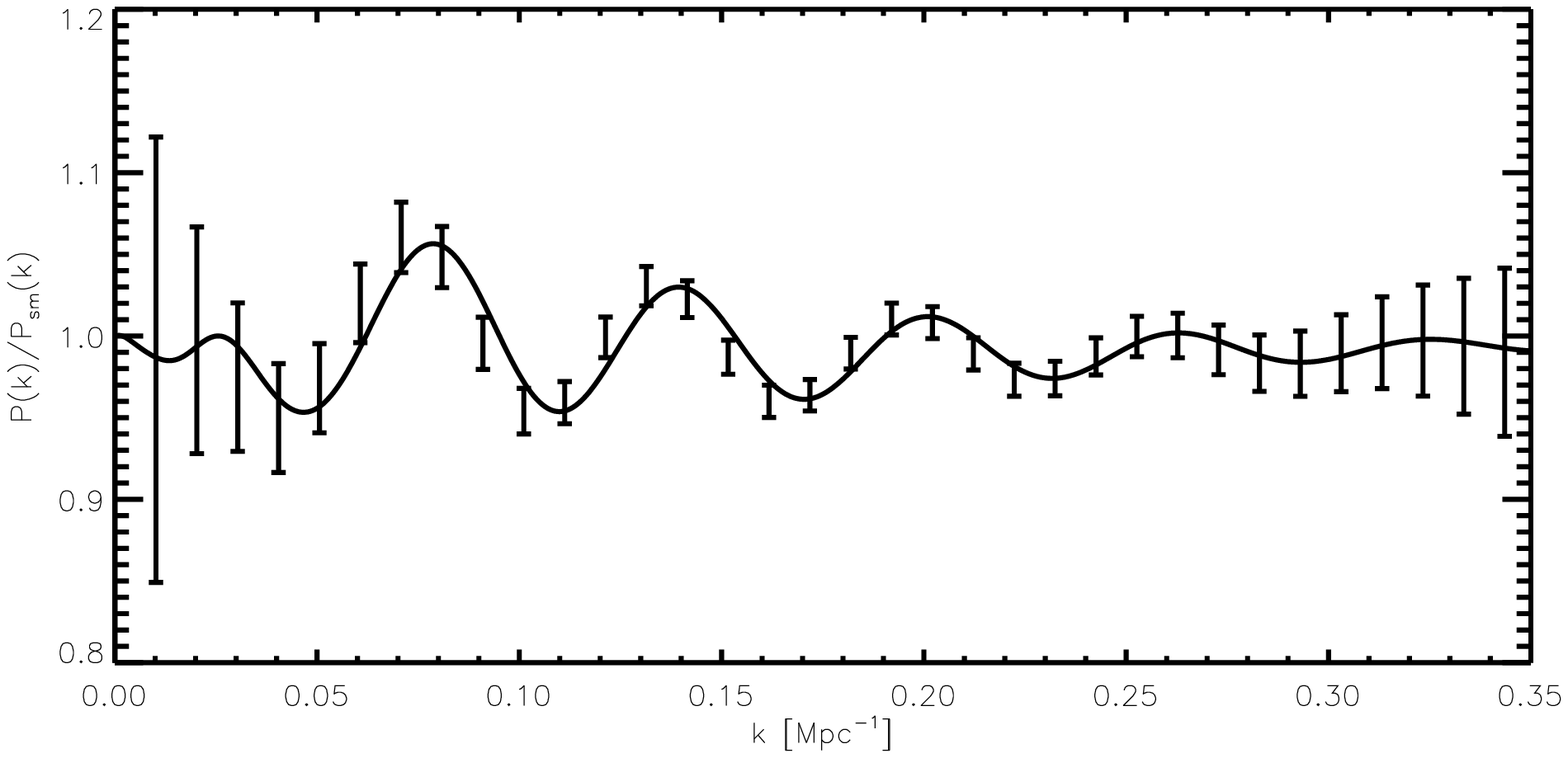}}
\caption{The measurement on the BAO of the matter power spectrum shown in flat $k$-space. {\it Left}: FAST experiment with frequency coverage $0.95$--$1.35$\,GHz; {\it Right}: BINGO experiment. The total signal-to-noise ratio with respect to the smoothed matter power spectrum is $10.5$ for extended FAST (left panel), $10.2$ for BINGO (right panel), and $8.1$ for original frequency range ($1.05$--$1.35$~GHz) in~\citet{Nan11}.} \label{fig:fbao}
\end{figure}

%
%
%

\begin{figure}[tbp]
\centerline{
\includegraphics[width=2.5in]{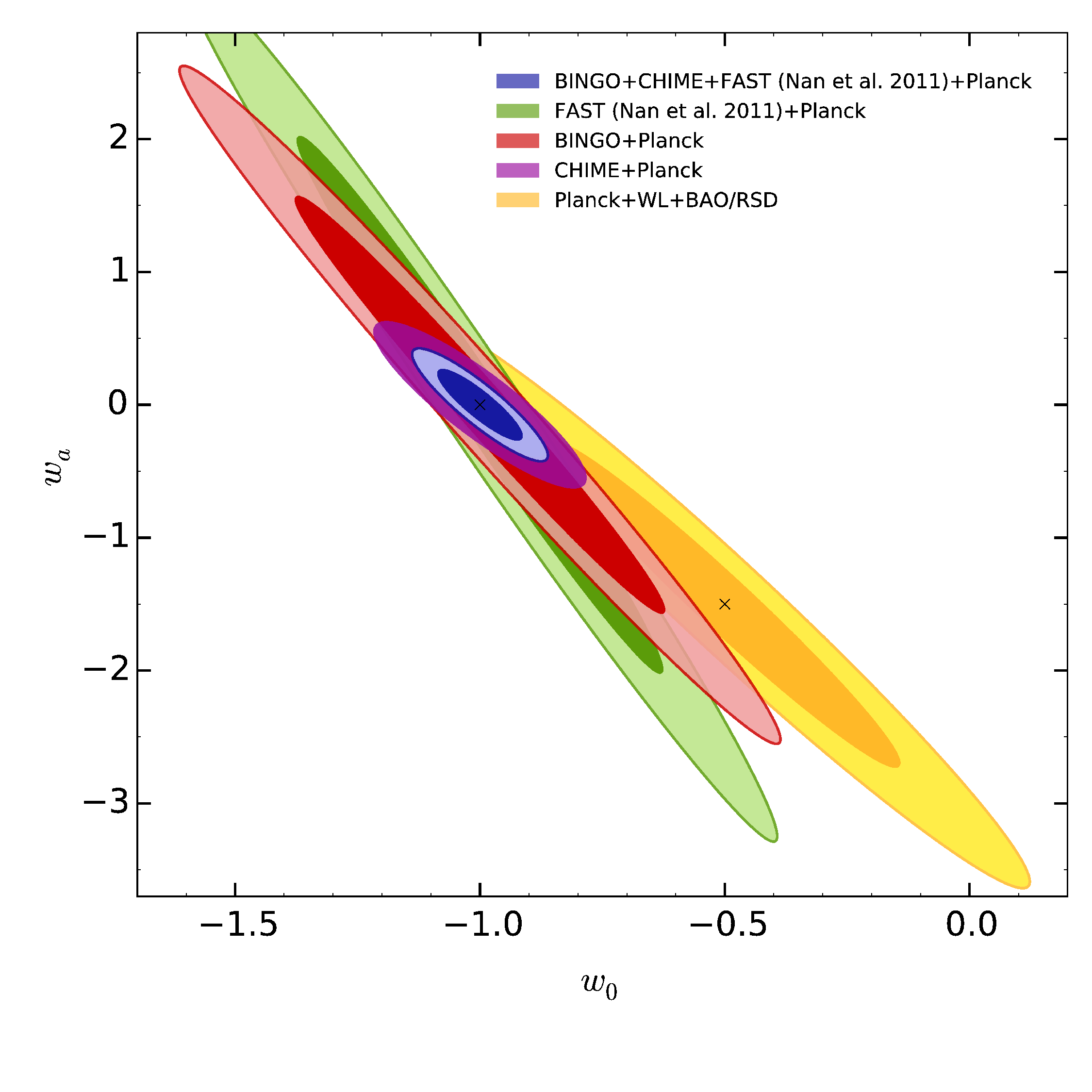}
\includegraphics[width=2.5in]{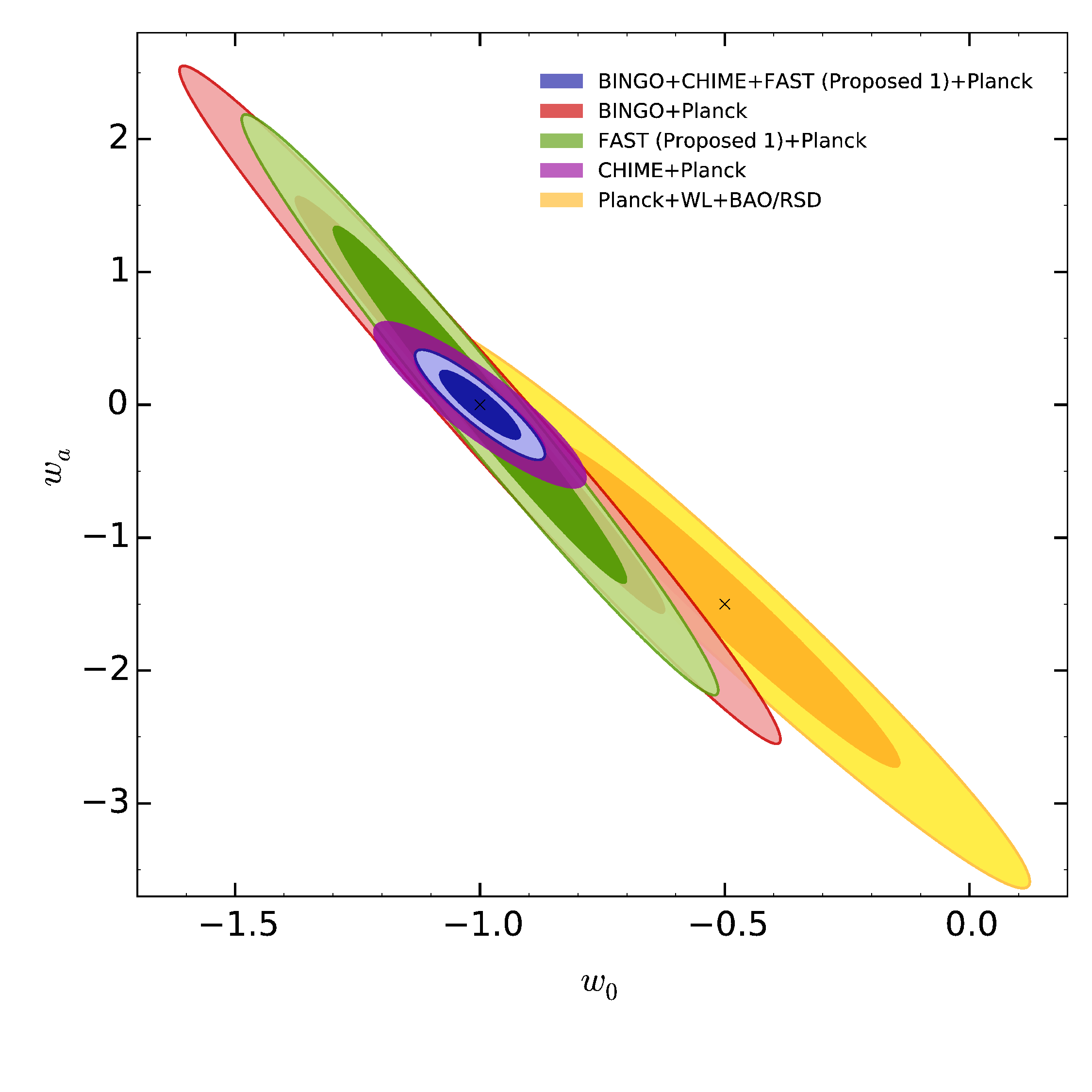}}
\caption{The joint forecasts on dark energy EoS parameter $(w_{0},w_{a})$, where $w(a)=w_{0}+(1-a)w_{a}$. The different prospective observations are shown in the colour contours. The difference between the left and right panels is the FAST scanning frequency band design. Left panel uses the design given in~\citet{Nan11}, and right panel uses the suggested design with a larger frequency coverage defined in Table~\ref{tab:surveys}. The yellow contour is the current constraint from~\citet{Planck-13}.} \label{fig:w0wa}
\end{figure}

\subsection{Constraints on BAO scales}
\label{sec:BAO-scales}

In Fig.~\ref{fig:fbao}, we plot the projected errorbars on the BAO wiggles of the matter power spectrum for FAST with the frequency coverage $0.95$--$1.35$\,GHz in the left panel, and for BINGO in the right panel. Comparing the two panels, FAST and BINGO provide similar constraints, and FAST is slightly better than BINGO on small scales (larger $k$-values) due to its better angular resolution. The total S/N (defined as (S/N)$_{\rm tot}=\left[\sum_{i} \left((P(k_{i})-P_{\rm smo}(k_{i}))/\sigma_{P(k_{i})} \right)^{2} \right]^{1/2}$ since errorbars at different $k$s are roughly uncorrelated) with respect to the smoothed matter power spectrum is $10.5$ for FAST, $10.2$ for BINGO, whereas it is $8.1$ for the original frequency range of FAST ($1.05$--$1.35$~GHz) in~\citet{Nan11}. However, we should bear in mind that BINGO is a targeted experiment designed for detecting the BAO in the distribution of HI, while FAST is a general purpose radio telescope with multiple science objectives. Therefore, it is unlikely that FAST will match the sensitivity of BINGO unless it conducts a full $1$-year observation of intensity mapping only.

\subsection{Constraints on Dark Energy}
\label{sec:de}

We further carry out the forecast on the projected constraints of dark energy EoS. The dark energy EoS is normally parametrized as $w(a)=w_{0}+(1-a)w_{a}$, where ($w_{0}$,$w_{a}$) represents the magnitude and first derivative of the dark energy EoS. The cosmological constant model corresponds to $(w_{0}$,$w_{a})=(-1,0)$. For doing this analysis, we use the the Fisher Matrix code of~\citet{Bull15} to calculate the likelihoods of cosmological parameters given various cosmological data. In Fig.~\ref{fig:w0wa}, we show the current tightest constraints from {\it Planck} 2015+Weak Lensing+RSD result in yellow contours~\citep{Planck-13} as a reference. We further combine BINGO, CHIME and FAST separately with {\it Planck} measurement. One can see that from the right panel that, if we use FAST with frequency coverage $0.95$--$1.35$\,GHz, it provides better constraints on $(w_{0},w_{a})$ parameter space than the original design in \citet{Nan11}. The marginalized constraints on $w_{0}$ for left and right panels are $w=-1.0 \pm 0.3$ and $w=-1.0 \pm 0.2$ ($1\sigma$ confidence level) respectively. Furthermore, by combining FAST, CHIME and BINGO with {\it Planck}, the contours are reduced. Thus, the tightest constraint in the future will be to combine {\it Planck} with all three HI intensity mapping experiments.

\section{Conclusion}
\label{sec:conclude}

In this proceeding, we have calculated how well the FAST telescope can measure the BAO scale and constrain the dark energy EoS. For one year integration with its 19 L-band receivers, FAST can provide the constraints of BAO wiggles similar to BINGO. So one needs a long integration time with FAST with the Focal Plane Array (FPA) to make a significant contribution even at low redshift. By extending the lower frequency limit of the 19 beam L-band receiver from $1.05$ to $0.95$ GHz, the total signal-to-noise ratio of BAO wiggles improves from $8.1$ to $10.5$. The dark energy EoS parameter is also better constrained. In this sense, combining the constraints with another receiver at higher redshift could improve the sensitivity of FAST depending on the number of beams of this receiver. 

The forecasts presented here have only considered the effects of thermal noise, and we assume no systematics and a perfect calibration of the data. There are a number of issues that must be considered carefully with radio telescope arrays such as the presence of $1/f$ noise, drifts in the gain, the effect of sidelobes, spillover, and RFI contamination. It needs to be demonstrated than the FAST experiment can control these effects down to the thermal noise level. In a more detail follow-up paper, we will generate realistic 21-cm maps for FAST including foreground emission and systematic effects by using the methods developed in~\citet{Bigot-Sazy15} and \citet{Olivari15}, and simulate some systematic effects. In that way, we will present for the first time the realistic simulation of FAST and obtain achievable and reliable cosmological constraints of FAST.

\acknowledgements 
The research leading to these results has received funding from the European Research Council under the European Union's Seventh Framework Programme (FP7/2007--2013) / ERC grant agreement no.~307209. CD also acknowledges support from an STFC Consolidated Grant (no.$\sim$ST/L000768/1). The authors acknowledge Phil Bull for making his code available.



\end{document}